# Analysing viewpoints in design through the argumentation process

**Géraldine Martin\*,\*\*; Francoise Détienne\*; Elisabeth Lavigne\*\***

\* INRIA-Eiffel Research Group "cognition and cooperation in design"
Domaine de Voluceau, Rocquencourt, BP 105, 78153 Le Chesnay, France

\*\*EADS AIRBUS-SA  BTE/SM/GDT-CAO M0101/9 316 route de Bayonne
31060 Toulouse cedex 03, France

**Abstract :** We present an empirical study aimed at analysing the use of viewpoints in an industrial Concurrent Engineering context. Our focus is on the viewpoints expressed in the argumentative process taking place in evaluation meetings. Our results show that arguments enabling a viewpoint or proposal to be defended are often characterized by the use of constraints. Firstly, we show that, even if some constraints are apparently identically used by the different specialists involved in meetings, various meanings and weightings are associated with these constraints by these different specialists. Secondly, we show that the implicit or explicit nature of constraints depends on several interlocutive factors. Thirdly, we show that an argument often covers not only one constraint but a network of constraints. The type of combination reflects viewpoints which have specific status in the meeting. Then, we will propose a first model of the dynamics of viewpoints confrontation/integration.

**Keywords :** Concurrent Engineering, viewpoint, constraints, design, assessment, argumentation

## 1  Purpose

In new design and production organizations, design is often the work of a multi-speciality, multi-location team, manoeuvring, according to the moment, with the same aim (co-design) or different aims (distributed design). In the collective design process, co-design phases are specifically devoted to the assessment of the global solution, integrating the solutions produced by the different designers at time t, or to the assessment, by his/her peers, of a solution produced by one designer at time t.

The aim of the study presented in this paper is to analyse the viewpoints brought into play in co-design. The chosen design context is a Concurrent Engineering process. This framework seemed to us to be the most relevant for studying the topic of « viewpoint », as the simultaneousness and confrontation of viewpoints during the development of the solution are assumed to be favoured by working in Concurrent Engineering (Darses, 1997).

Aerospatiale Matra Airbus has conducted the re-engineering of its design processes in a Concurrent Engineering procedure, in order to better master costs, schedules and quality in the design of its products. This industrial development is assisted by cognitive ergonomics research work, which is the framework of this study. We are analysing this setting up of a Concurrent Engineering methodology. A previous study focused on the coordination processes in distributed design (Martin, Détienne &Lavigne,1999). The industrial aim is to derive ergonomic recommendations at software level (digital mock-up, technical database) and organizational level (meeting methodology, definition of roles) in order to assist the confrontation and integration of viewpoints in multi-speciality design.

After a brief presentation of our theoretical framework and working hypotheses, we present an empirical study aimed at analysing the use of viewpoints in an industrial Concurrent Engineering context. Our approach is strongly oriented by cognitive ergonomics work on the notion of constraint, and linguistics work on argumentation.

## 2 Theoretical framework and working assumptions

The confrontation of knowledge and the integration of viewpoints is at the heart of the cooperative mechanisms implemented in co-design. A new research topic is to characterize the viewpoints of the various players involved in collective design (designers themselves, and production and maintenance specialities) and the cooperative modes that enable these different viewpoints to be integrated.

During the design process, different viewpoints are implemented. On the basis of the work performed in different disciplines - Artificial Intelligence (Wenger, 1987), cognitive ergonomics (Rasmussen, 1979; Darses, 1997), ethnomethodology (Bucciarelli, 1998), Computer-Supported- Cooperative Work (Schmidt, 1994), an initial general definition of the notion of "viewpoint" would be : " for a person, a particular, personal, representation of an object to be designed". We are now going to develop this definition a little more precisely.

In the representation of the object to be designed, and also of its design, design constraints seem to us to play a predominant part. For design problems, the solutions are not unique and correct but various, and more or less satisfactory according to the constraints that are considered. The designers assess the solutions they develop according to their own specific constraints, which reflect their own specific viewpoints, in relation with the specificity of the tasks they perform and their personal preferences (Eastman,1969; Falzon et al,1990).

Constraints are cognitive invariants which intervene during the design process. The notion of constraints has been understood from different angles (1) according to their origin - prescribed constraints, constructed constraints, deduced constraints, (2) according to their level of abstraction, and (3) according to their importance – validity constraints and preference constraints (Bonnardel,1999).

The presentation of the object to be designed is characterized according to an abstract-concrete line or abstraction hierarchy (Rasmussen,1979). The different levels of abstraction are integrated into each state of the solution. This can reflect functional, structural or physical viewpoints all along the design process (Darses,1997; Darses & Sauvagnac, 1997). Factors such as the field of expertise and specific technical interest play a role in this representation. Indeed, several participants see the design object differently according to the specificities and constraints specific to their speciality. In addition, for the same speciality, the representation will be variable according to the problem to be solved.

To conclude, our definition is that a viewpoint is characterized by the implementation of a certain combination of constraints that are specific to each speciality and dependent on the problem to be solved.

Our working assumption is that viewpoints are expressed, more or less explicitly, in multi-speciality meetings, aimed at co-design, in particular, the assessment of solutions. It is thus on the analysis of these meetings that we have focussed our empirical work. In design activities, the assessment intervenes (1) to appreciate the suitability of partial solutions to the usual state of resolution of the problem, and (2) to select one of the solutions envisaged (Bonnardel,1999). It is in assessment meetings that we should observe the confrontation of the viewpoints of the various participants in design.

Owing to the collective nature of the activity, viewpoints should be expressed, more or less explicitly, through argumentation (Plantin,1996). In the argumentative dialogue, a proposer will express a viewpoint that will be argued about by presenting a certain amount of information substantiating the initial proposal.

## 3 Methodology

### 3.1 Context

We conducted this study during the definition phase of an aeronautical design project, lasting three years, in which the participants work in Concurrent Engineering to design the centre section of an aircraft. These participants use Computer Assisted Design (CAD) tools and a technical Data Management System (PDM). About 400 people with 10 different specialities are involved. These specialities are the traditional design office specialities (structure, system installation, stressing), specialities that used to intervene further downstream (maintainability, production) and new specialities that have appeared with the introduction of CAD and PDM tools.

### 3.2 First phase

#### 3.2.1 Collection of data during meetings

All the specialities work on the same part of the aircraft but each person according to his technical competence. "Informal" inter-speciality meetings are organized, as needed, to assess the integration of the

solutions of each speciality into a global solution. We took part in 7 of these meetings as observers:
- Five meetings involved upstream design office players (designers from structure and systems installation specialits);
- Two meetings involved upstream-design office and dowstream players (from production or maintenance specialities).

On the basis of audio recordings and notes taken during the meeting, we retranscribed the full content of the meetings. Each meeting involved 3 to 6 players.

### 3.2.2 Coding scheme

The protocols resulting from the retranscriptions were broken down according to the change of locuters. Each individual participant statements correspond to a "turn". Each turn was coded according to the following coding scheme and broken down again as required to code finer units. Our coding scheme comprises two levels, a functional level and an argumentative level.

The functional level highlights the way in which collective design is performed. Each unit is coded by a mode (request/assertion), an action (e.g., assess) and an object (e.g., solution n). At this level, a turn can be broken down into finer units according to whether there is a change in mode, activity or object.

The argumentative level brings out the structure of the speech on the basis of a dialogue situation. We coded the proposals for solutions made and the different types of arguments used by the speakers during the meetings. Functional units (but not all as some units clearly do not belong to this process) were assigned three kinds of role in the argumentative process:
- Proposal X: solution X is proposed by one or several participants;
- Argument +(X) or – (X): arguments supporting or not supporting the proposal are advanced by the participants;
- Resolution: the proposal is accepted or rejected by all the participants or there is an absence of conclusion

The nature of the arguments was further refined. In particular we examined whether:
- one or several constraints were used in the argument;
- an example was brought out to convince the others (argument by example used in analogical evaluation);
- the argument had the status of argument of authority: In this case, an argument is presented as inconstestable and therefore it has a particularly strong weight in the negotiation process. An argument can take the status of argument of authority depending on :
- the status, recognised in the organisation, of the speciality that expresses it.
- the expertise of the proposer.
- the "shared" nature of the knowledge to which it refers.

Furthermore we detected the converging moves (agreement between participants on the acceptance or rejection of a proposal) and diverging moves .

## 3.3 Second phase

### 3.3.1 Auto-confrontations with coded protocols of meetings

We conducted interviews afterwards with the various participants of meetings to validate the coding we had made and make explicit a certain amount of information that was implicit in the meetings. As our focus was on the analysis of viewpoints through the arguments expressed during evaluation meetings, in particular through the notion of constraints, our primary concern was to validate our coding of the argumentation process.

We gave to each participant our coding of the meeting(s) where he/she took part, and ask him/her to assess our coding and to make explicit the case where one or several constraints were implicit in an expressed argument but in fact founded the argument itself. This allowed us to make appear, in the argumentation process, the distinction between:
- Argument with explicit constraint(s): e.g "If we have a 160mm pulley, and we've only got 140, were going to have a problem " (explicit system-installation constraint)
- Argument with implicit constraint(s) : e.g " this fractured on the other aircraft" (the implicit constraints are stress and structure)

### 3.3.2 Tests with constraints

Our second concern was to identify what representation each specialist had about constraints: in particular the representation of the meaning assigned to a constraint expressed a certain way and the ordering between constraints. Based on previous work, we thought that these representations may depend on the expertise of the players, in particular his/her speciality, but also on the context (the problem-situation addressed). Thus, our tests were constructed depending on the problem-situation, i.e. the meeting, in which constraints had been used.

For each meeting, we collected the constraints used (either explicitly or implicitly) and presented

the list to each participant of this meeting. Our question concerned:
- for each constraint: to give their meaning;
- for all constraints: to order them as a function of their importance in this design-problem-situation.

# 4 Results

The first type of result involved the way in which the proposals for solutions are assessed during these meetings. We have revealed the existence of assessment modes in these meetings as well as their combination (Martin, Détienne & Lavigne, 2000):
- analytical assessment modes: systematic assessment of proposal according to constraints;
- comparative assessment modes: systematic comparison between alternative proposals according to constraints;
- analogical assessment modes : transfer of the result of the assessment of an analogical solution (source) to the current proposal (target).

In this paper, our focus is on the viewpoints expressed in the argumentative process. We will present four types of result concerning the constraints expressed by the different specialities. Indeed, arguments enabling a viewpoint or proposal to be defended are often characterized by the use of constraints.

Firstly, we will show that, even if some constraints are apparently identically used by the different specialists involved in meetings, various meanings and weighting are associated with these constraints by these different specialists.

Secondly, we will show that, in the argumentation process, constraints can be explicit or implicit in the argument as it is expressed by a speaker. The implicit or explicit nature depends on several interlocutive factors.

Thirdly, we will show that an argument often covers not only one constraint but a network of constraints. The type of combination reflects viewpoints which have specific status in the meeting.

Then, we will propose a first model of the dynamics of viewpoints confrontation/integration.

## 4.1 Meaning and weighting of constraints

Constraints used in the argumentation process are of two kinds:
- Prescribed constraints independent of the speciality (or skill): those constraints are prescribed in the design specification and, a priori, shared by all the players of the design process;
- Derived constraints specific to a speciality.

We found that, even though some constraints used by different players in a meeting are the same at a surface level (same terminology), these constraints may have different meanings in the viewpoints expressed by players from different specialities. Also, the level of refinement selected may be different according to the speciality.

### 4.1.1 Selection of a meaning for a skill-independent-constraint

We observed that the same constraint (the same terms are used by different players in a meeting) can have different meanings according to the speaker's speciality.

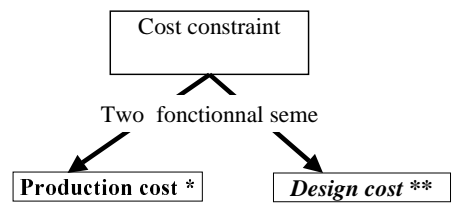

\* **Meanings according Design Office field 1**
\*\* *Meaning according Design Office field 2*

**Figure 1 :** Selection of a meaning for a skill-independent-constraint

In this case it is necessary to distinguish the two slopes of the sign, the signifier and the meaning. The meaning can have the same generic seme for different speakers but a very different functional seme. Figure 1 illustrates that a cost constraint can, for one speciality, mean "production cost" and, for another speciality, mean "design cost". It seems particularly true for general constraints prescribed for all the players of the design process (e.g., the cost) as opposed to constraints derived by a speciality (e.g., structure).

### 4.1.2 Selection of a refinement level in a hierarchical network of a skill-dependent-constraint

We found that some constraints expressed in the argumentation process may be organized hierarchically along different levels of refinement. For example, a maintenance constraint may be refined as three constraints: accessibility constraint, dismounting constraint and mounting constraint. However, when we analysed the skill-dependent constraints used for expressing the viewpoints of different players, we identified some gaps between the level of refinement selected and used in the argumentation process according to the speaker's speciality. For a constraint specific to a skill, the level of refinement is more detailed for the speciality which represents this skill and more general for the other speciality. Two examples are given in Figure2.

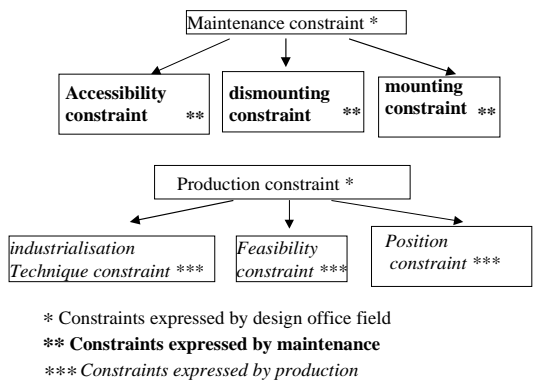

\* Constraints expressed by design office field
\*\* **Constraints expressed by maintenance**
\*\*\* *Constraints expressed by production*

**Figure 2 :** Selection of a level of detail in the hierarchical network of constraints specific to a skill

### 4.1.3 Constraints weighting

Constraints used and their weighting, which also founds the viewpoint of the participants, depend on several factors.
- The participant's speciality;
- The interlocutors;
- The design-problem situation.

The selection of constraints depends on speaker speciality and on the interlocutors. In general, constraints taken into account in a particular meeting are those constraints specific to the specialities involved in the meeting in addition to the prescribed constraints. However skill-dependent-constraint weighting depends on speaker speciality. Whereas we found a high intra-speciality agreement on constraint weighting, we found disagreement between specialities. An example is given in Table1.

| | structure/ hydraulics Meeting | |
|---|---|---|
| | *Hydraulics specialists* | **Structure specialists** |
| Level 1 | *System installation production time* | Maintainability |
| Level 2 | Maintainability | **Structure Stress** |
| Level 3 | Growth of problem | production |
| Level 4 | *Frontier* | System Installation |
| Level 5 | **Structure Stress** | |

**Table 1:** Constraints ranking from the most important (level 1) to least important (level 4) for two specialities

The constraints which are specific to Hydraulic system intallation specialists (in blod) are : system installation and frontier. The constraints which are specific to Structure specialists (initalic) are : structure and stress. We can see that, even if most of these constraints are used by the two specialities involved in the meeting, the way each speciality orders those constraints by importance is different. Each specialist ranks his/her own constraints as more important than the constraints of his/her interlocutors.

Furthermore, we can see in this example, that some constraints are used only by one speciality: time, growth of problem, frontier are used only by hydraulics specialists.

Constraints weighting also depends on the problem in hand. For example, we observed for the same speciality, air system installation, that constraint weighting varied between two problems processed sequentially in a meeting: the maintainability constraint was ranked 3 for problem A and 1 for problem B. Furthermore the production constraint was evoked only for problem A.

## 4.2 Mechanisms of constraint clarification

Constraints can be explicit or implicit in the argument as it is expressed by a speaker. The mechanism of constraint clarification may depend on several interlocutive factors. Of course, it may depend on an explanation request made by another participant of the meeting: this is a rather straightforward mechanism that we observed in a systematic way. It may also depend on the postulate of shared knowledge made by the speaker: this factor was difficult to assess.

We have identified two other conditions of constraint clarification : diverging move between specialities; reinforcement of intra-speciality-consensus.

| Structure/ system installation meeting | | | | |
|---|---|---|---|---|
| Turn | spec | Arg | Constraints | Diverging/ converging moves |
| 75 | IS | Pro | | |
| … | … | … | | |
| 78 | St1 | 14 | Stress **imp** /structure **imp** | |
| 79 | St2 | … | | **Imp divergence** |
| 80 | St1 | 15 | Stress exp / structure **imp** | |
| 81 | St2 | 16 | Stress exp /structure **imp** | |
| 82 | IS | | | |
| 83 | IS | | | Exp divergence |

**Legend** : St(X) Designer -Struc X, IS : Designer-IS, Prop : proposal, Arg : Argument, imp : implicit, Exp : explicit Blod : constraint clarification following divergence; White: absence of argument or constraint

**Table 2 :** Implicit divergence leading to constraint clarification

Table 2 gives an example of an implicit divergence which leads a participant to make explicit constraints he used in previous arguments. This table highlights the chronological sequence of the situation at a structure/system installation meeting. The Structure-designer (St1) put forward arguments for rejecting a solution which was proposed previously by a system installation specialist. St1 does so by referring to a similar problem, saying:

*"This fractured on the other aircraft* [Arg 14 Stress/structure constraints], *".*

Even if this argument is founded on two constraints, stress and structure, these constraints remain implicit in what is said by the designer.

Faced with the lack of reaction from the IS-designer, St1 argues still further:
"Why? Because according to the computation there was a relative displacement of the beam of approximately 2mm with respect to the other one [Arg 15 Stress/structure constraints]"

The lack of reaction from the IS-designer led the Struc-designers to assume an implicit divergence on the part of the IS-designer. He thus felt compelled to argue his rejection of the solution by making the stress constraint explicit. This divergence is, moreover, explained just after the 82$^{nd}$ successive contribution to the discussions, with the following words:
"So we contacted several people dealing with the electrical installation on the other aircraft, and had no feedback of any incidents at that level".

| structure/Installation System meeting | | | |
|---|---|---|---|
| Turn | speciality | Arg | Constraints |
| 78 | St1 | 3 | Time explicit /cost implicit/ program-Study implicit |
| 80 | St1 | 4 | Stress explicit |
| 81 | St2 | 5 | Time explicit /cost implicit/ **program-Study explicit** |
| 81 | St2 | 6 | Stress explicit / structure implicit |

**Legend** *: St(X) Designer-Struc X; in Red: clarification of constraint for reinforcement of consensus, Arg : argument.*

**Table 3** : Reinforcement of consensus by expliciting constraints

Table 3 shows, in chronological order a strengthening of the consensus of opinion by another representative from the same skill. In this structure/ system installation meeting, two Struc-Designers – St(1) and St(2) - are present, and try to convince the system installation designers of their point of view.

In this System structure/installation meeting, the two Struc-Designers reject a solution proposed by the IS-Designers. To show his disagreement with the IS-designers, St(1) puts forward his arguments for rejecting the solution (78$^{th}$ contribution to the discussions):

"*Because in that case they would have to do another study [Arg 3 constraints relating to program and study deadlines and costs], and add material [Arg 4 cal-constraint and struc-constraint]*".

St(2) goes even further than St(1) by explaining the design constraint (program-study constraint) left implicit by St(1).

He says *"The complete study already conducted will have to be done again [Arg 5 constraints relating to program and study deadlines and costs], as there is an offset of the box beam [Arg 6 cal-constraint and struc-constraint].*

By using this process for strengthening the consensus of opinion, St(2) backs up what St(1) has already said. He emphasizes this mechanism by using two arguments (arg5 and arg6) that refer to the same constraints used by St(1) in arguments 3 and 4. By doing this he obliges the IS-Designers to justify the advantages of the solution they propose even further. By using these means, the Struc-Designers endeavour to impose their viewpoint on the solution.

## 4.3 Constraint combination

As we have already seen in the examples above, a viewpoint is most of the time founded on a combination of constraints. These combinations are of three types and have different status as viewpoints:

- Combination of skill-independent-constraints: it represents a <u>shared viewpoint</u> as these constraints are prescribed and taken into account by all the players. However, we should not forget that this apparently shared representation may hide some gap between the meaning that each speciality associates with these constraints;
- Combination of skill-independent-constraints and own skill-dependant-constraints: it represents a <u>speciality-specific viewpoint</u>. Two variants were observed:
    - During the argumentation process: Players of speciality-1 may use this kind of combination with skill1-dependent-constraints they consider less important: it is a "weak" speciality-specific viewpoint. We argue that this combination is a way to have one's own point of view accepted by the interlocutors. Indeed, associating one's own constraints, in particular those weighted as less important, with prescribed constraints accepted (and not contestable) by the various specialists is a way to make a stronger argument in the argumentation process.
    - At the end of the argumentation process: Players of speciality-1 use this kind of combination with skill1-dependant-constraints they consider more important: it is a "strong" speciality-specific viewpoint. It is a way to check that constraints are satisfied as a result of the negotiation, in particular the prescribed ones and one's own constraint weighted more important.
- Combination of skill1-dependant-constraints and skill2-dependant-constraints : it represents an <u>integration of viewpoints</u>. However we should not forget that the same combination may be weighted differently by players of specialisty-1 and players of specialisty-2.

## 4.4 Dynamics of viewpoints confrontation/integration

We found that the three types of viewpoints were introduced in an invariant order in the argumentation process. Furthermore, there are related to certain kinds of evaluation. Figure 3 shows the dynamics of viewpoints confrontation/integration.

In a first step, each specialist evaluates the proposed solution with an analytical assessment mode. This is done on the basis of a combination of skill independent constraints or a combination of one's own skill-specific constraints. In this way, <u>shared viewpoints</u> or <u>speciality-specific viewpoints</u> are expressed. If this process does not allow the various players to converge, which is generally the case, then a second step occurs.

In the second step, each specialist evaluates the proposed solution with an analytical /comparative assessment mode or analytical /analogical assessment mode:

- This is done on the basis of a combination of skill independent constraints and one's own skill-specific constraints. In this way, a <u>speciality-specific viewpoints</u> is presented. Or

- Or this is done on the basis of a combination of skill1 and skill2-specific constraints. In this way, an <u>integrated viewpoint</u> is found. This is based on shared knowledge concerning the evaluation of the source or alternative solution. We found that two mechanisms may be involved in this integration of viewpoint. One mechanism is to explicitly negotiate constraints when the search for alternative solutions is in an impasse. Another mechanism is to evoke shared knowledge concerning the evaluation of a source (previous solution developed for an analogous problem) for which an integrated viewpoint was found.

If this process does not allow various players to converge, then a third step occurs.

In the third step, argument of authority are used and allow a skill player to "impose" one's own point of view.

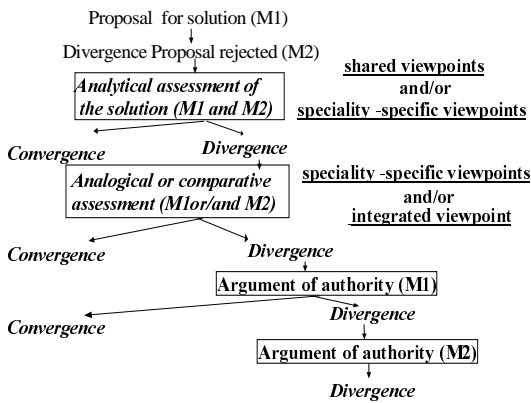

**Figure 3** : Dynamics of viewpoints confrontation/integration

# 5  Further work

This paper presents a first attempt to analyse viewpoints involved in design within an ergonomic theoretical framework. Viewpoints have been analysed through the argumentation process. To go further we plan to examine deeper the following issues:
- Characterizing viewpoints by a certain level of abstraction, i.e. functional, structural or physical;
- The relationship between constraint meaning/weighting and viewpoints in the argumentation process;
- The assessment of the argumentation process: which are the conditions leading to a convergence between participants, in particular, to an integration of viewpoint?

Our results can be a basis to specify meeting methodology and support for meetings such as argumentative system (see for example, Lonchamp, 2000). For example, we believe that it is important to support in some way the distinct assessment modes. Furthermore, viewpoints, as founded by combination and weighting of constraints, could be represented in some way both for supporting the argumentative process and for ensuring the traceability of design decisions.